# ARPM-net: A novel CNN-based adversarial method with Markov Random Field enhancement for prostate and organs at risk segmentation in pelvic CT images

Zhuangzhuang Zhang

*Department of Computer Science and Engineering, Washington University, One Brookings Drive, Campus Box 1045, St. Louis, Missouri 63130*

Tianyu Zhao

*Department of Radiation Oncology, Washington University School of Medicine, 4921 Parkview Place, Campus Box 8224, St. Louis, Missouri 63110*

Hiram Gay

*Department of Radiation Oncology, Washington University School of Medicine, 4921 Parkview Place, Campus Box 8224, St. Louis, Missouri 63110*

Weixiong Zhang [a]

*Department of Computer Science and Engineering, Department of Genetics, Washington University, One Brookings Drive, Campus Box 1045, St. Louis, Missouri 63130*

Baozhou Sun [a]

*Department of Radiation Oncology, Washington University School of Medicine, 4921 Parkview Place, Campus Box 8224, St. Louis, Missouri 63110*

a) Correspondence and requests for materials should be addressed to W.Z (weixiong.zhang@wustl.edu) or B.S (Baozhou.sun@wustl.edu)

**Purpose:** The research is to develop a novel CNN-based adversarial deep learning method to improve and expedite the multi-organ semantic segmentation of CT images and to generate accurate contours on pelvic CT images.

**Methods:** Planning CT and structure datasets for 120 patients with intact prostate cancer were retrospectively selected and divided for 10-fold cross-validation. The proposed adversarial multi-residual multi-scale pooling Markov Random Field (MRF) enhanced network (ARPM-net) implements an adversarial training scheme. A segmentation network and a discriminator network were trained jointly, and only the segmentation network was used for prediction. The segmentation network integrates a newly designed MRF block into a variation of multi-residual U-net. The discriminator takes the product of the original CT and the prediction/ground-truth as input and classifies the input into fake/real. The segmentation network and discriminator network can be trained jointly as a whole, or the discriminator can be used for fine-tuning after the segmentation network is coarsely trained. Multi-scale pooling layers were introduced to preserve spatial resolution during pooling using less memory compared to atrous convolution layers. An adaptive loss function was proposed to enhance the training on small or low contrast organs. The accuracy of modeled contours was measured with the Dice similarity coefficient (DSC), Average Hausdorff Distance (AHD), Average Surface Hausdorff Distance (ASHD), and relative Volume Difference (VD) using clinical contours as references to the ground-truth. The proposed ARPM-net method was compared to several state-of-the-art deep learning methods.

**Results:** ARPM-net outperformed several existing deep learning approaches and MRF methods and achieved state-of-the-art performance on a testing dataset. On the test set with 20 cases, the average DSC on the prostate, bladder, rectum, left-femur, and right-femur were 0.88($\pm$0.11), 0.97($\pm$0.07), 0.86($\pm$0.12), 0.97($\pm$0.01), and 0.97($\pm$0.01), respectively. The Average HD (mm) on these organs were 1.58($\pm$1.77), 1.91($\pm$1.29), 3.14($\pm$2.39), 1.76($\pm$1.57), and 1.92($\pm$1.01). The Average Surface HD (mm) on these organs are 2.11($\pm$2.03), 2.36($\pm$2.43), 3.05($\pm$2.11), 1.99($\pm$1.66), and 2.00($\pm$2.07).

**Conclusion:** ARPM-net was designed for the automatic segmentation of pelvic CT images. With adversarial fine-tuning, ARPM-net produces state-of-the-art accurate contouring of multiple organs on CT images and has the potential to facilitate routine pelvic cancer radiation therapy planning process.

## 1. INTRODUCTION

Radiotherapy treatment planning requires accurate contours for maximizing target coverage while minimizing the toxicities to the surrounding organs at risk (OARs).[1] The diverse expertise and experience levels of physicians introduce large intraobserver variations in manual contouring.[2] Interobserver and intraobserver variation of delineation results in uncertainty in treatment planning, which could compromise treatment outcome.[3] Manual contouring by physicians in current clinical practice is time-consuming, which is incapable of supporting adaptive treatment when the patient is on the couch. Accurate segmentation of prostate and OARs is crucial in pelvic CT images.[4] Pelvic CT image segmentation is challenging because the prostate has fuzzy boundaries with the background due to low contrast and variations in organ size, shape, and intensity.[4]

Many algorithms for automated segmentation have been proposed to improve the accuracy, reliability, and efficiency of delineation in pelvic CT images. Matinez *et al* used a geometrical shape model tuned by a multi-scale edge detector for pelvic structure segmentation and achieved the Dice similarity coefficient (DSC) of 0.91 for prostate, 0.94 for bladder, and 0.89 for rectum.[5] Macomber *et al* used deep decision forests of radiomic features for auto-segmentation of prostate anatomy and with the DSC of 0.94-0.97 for bladder, 0.96-0.97 for femurs, 0.75-0.76 for prostate, 0.71-0.82 for rectum, and 0.49-0.70 for seminal vesicles.[6] Gao *et al* used regression-based deformable models and multi-task random forest and achieved the DSC of 0.88 for prostate, 0.86 for bladder, and 0.85 for rectum.[7] With the development of Convolutional Neural Network (CNN), deep learning models have outperformed the state-of-the-art methods in many visual recognition tasks.[8] Cha *et al* developed a computerized system for bladder segmentation in CT urography (CTU), in which CNN was utilized to distinguish the inside and the outside of the bladder using 160,000 regions of interest (ROI).[9] Fully Convolutional Network (FCN) extended traditional CNN models designed for classification tasks by replacing fully-connected classifying layers with fully convolutional layers, which enabled the model to perform semantic pixel labeling.[10] Ma *et al* proposed to use a 2D FCN for initial prostate segmentation and a multi-atlas label fusion for label refinement on CT images.[11] The combination of the two methods achieved the DSC of 0.87 on the prostate.[11] Wang *et al* proposed a 3D deeply supervised dilated FCN on CT images and achieved the DSC of 0.85 ± 0.04 for prostate evaluated with 15 pelvic CT images.[12]

U-net was designed as an upgraded version of FCN, which made the up-sampling path trainable by an encoder-decoder architecture.[8,10] Forward passing of feature maps provided a larger context that guided the up-sampling process of both FCN and U-net.[8] Several CNN-based U-net style models have been proposed for medical image semantic segmentation since U-net was proposed. Ma *et al* developed a U-Net based deep learning approach (U-DL) for bladder segmentation in CTU, and they used both 2D slices and 3D CT volumes as input. The U-DL provided more accurate bladder segmentation than deep learning convolution neural network with level sets.[13]

Kazemifar *et al* used a 2D U-net structure to perform auto-segmentation of the ROI on pelvic CT scans.[14] They later developed a multi-channel 2D U-Net followed by a 3D U-Net and modified the encoding arm with aggregated residual networks.[4] Combining the localization network and segmentation network improved the DSC result on the prostate from 0.88 ± 0.12 to 0.90 ± 0.02. The implementation of two U-net style neural networks increased the memory requirement for training the model.[4]

Considering the trade-off between localization accuracy and classification performance of Deep Convolutional Neural Networks (DCNNs), Markov Random Field (MRF) has been often used as a post-processor to perform label refinement.[15] DeepLabV2 used a fully-connected MRF to generate smoother label boundaries.[16] Deep Parsing Network (DPN) integrated MRF into the last few layers of CNN and trained them jointly.[17] Generative Adversarial Network (GAN) was originally proposed to implicitly represent high-dimensional probability distributions and generate style-wise plausible datapoint.[18] Employing an adversarial training scheme can provide a trustworthy discriminator to separate predictions from ground-truth labels.[19] Adversarial networks for image semantic segmentation performed well on the Pascal VOC 2012 dataset and the Stanford Background dataset.[20]

In this study, we proposed a novel adversarial multi-residual and multi-scale pooling MRF-enhanced network (ARPM-net) approach for semantic segmentation of multi-organs in CT images. Taking advantage of the well-established U-net method, we introduced several innovative ideas in designing the architecture of the new method, including a novel adaptive loss function for dealing with diverse sizes and variant contrasts of multiple organs, a novel adversarial training scheme for improving model quality, multiple residual and multiple scale pooling operations for fast feature extraction, and contour enhancement using MRF modeling. The design and development of ARPM-net hinged up two hypotheses. First, deep-learning techniques are better suited for dealing with many technical issues, e.g., diverse organ sizes, variable contrast, and image brightness on medical images, and can deliver better performance, i.e., accuracy and speed, than the conventional methods for medical image processing. Second, while deep learning has already been adopted for medical data, e.g., 2D U-net[8] and Multi-residual U-net[21], they can be further improved by more advanced deep-learning techniques, specifically adversarial neural networks[18], for generating more accurate contours on small and low-contrast organs.[22]

## 2. MATERIALS AND METHODS

### 2.1 Dataset

Institutional review board (IRB) approval was obtained for this study from Washington University of St. Louis. Planning CT and structure data sets for 120 intact prostate cancer patients were retrospectively selected. We split the 120 cases into a set of 100 cases for training and a set of 20 cases for testing for 10-fold cross-validation and testing. For cross-validation, 100 cases were divided for 10-fold cross-validation, in which 90 cases were used for training, and 10 cases for validation each time. We tested our hypotheses based on the performance of 10-fold cross-validation, and we used the performance on 20 test cases for comparing the new model with other state-of-the-art methods. All CT images were acquired using a 16-slice CT scanner with an 85 cm bore size (Philips Brilliance Big Bore, Cleveland, OH, US). The target organ was the prostate, and OARs included the rectum, bladder, left femur, and right femur. The CT images were acquired with a $512 \times 512$ matrix and 1.5 mm slice thickness. Each patient has 150-200 slices. The manually delineated contours served as ground truth in this study. The OARs were drawn by the same imaging technician, and the prostate contours were drawn by two radiation oncologists with over ten years' experience treating prostate cancer and a consensus contour generated using the Eclipse treatment planning system (Varian Medical Systems, CA). We fused the CT with MRT, which allowed for accurate delineation of the prostate. Individual lesions were not contoured as the standard clinical practice is to treat the entire prostate. Despite lesions that can sometimes be visualized in the MRI, prostate cancer tends to have a multifocal nature, and treating the entire gland is standard. Patients who had a prostatectomy were excluded.

## 2.2 Network Architecture

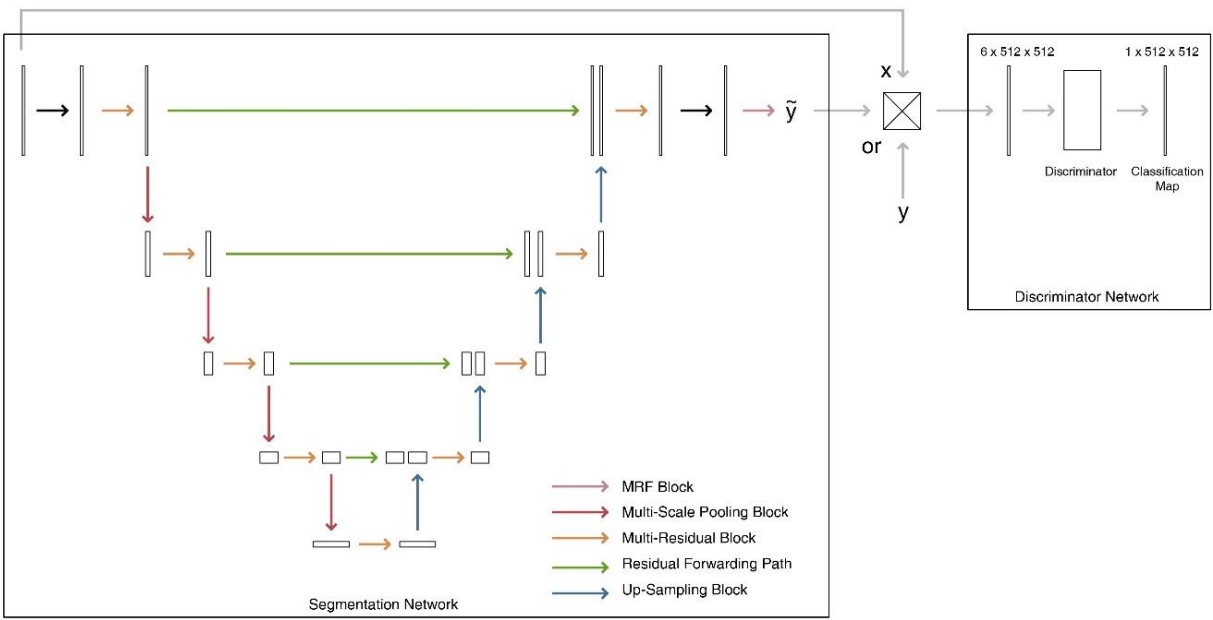

Figure 1. The ARPM-net Structure, where $x$ is the original CT, $\tilde{y}$ the one-hot encoding of a label prediction, and $y$ the one-hot encoding of the ground-truth label. Either $y$ or $\tilde{y}$ is used to calculate the product with $x$.

The ARPM-net has two major components (Figure 1), a segmentation network (S-net) and a discriminator network (D-net). Both components were used during training, but only the S-net was applied for prediction and evaluation.

Following the U-net structure[8], the S-net consists of a down-sampling encoder, an up-sampling decoder, and forwarding paths. It has the following major building blocks:

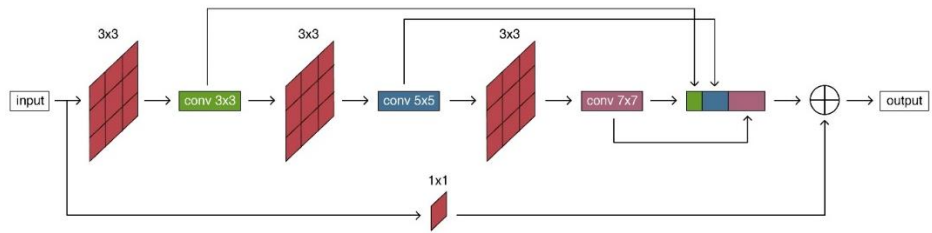

Figure 2. The multi-residual Block

- The Multi-Residual Block (Figure 2) is adapted from the Multi-residual U-net, in which 5x5 and 7x7 filters are factorized as a succession of 3x3 filters.[21] An increasing number of filters are also used in three successive layers gradually. The ratio of channels in conv3x3, conv5x5, conv7x7 is 3:5:8. A residual connection is employed after each 3x3 convolution layer. The 1x1 filter is used to conserve dimensions.

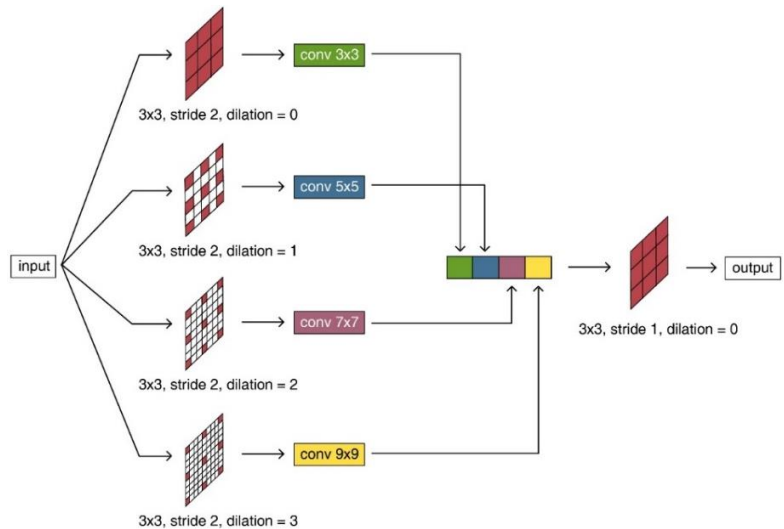

Figure 3. The multi-scale Pooling Block

- The Multi-Scale Pooling Block (Figure 3) pushes the feature maps through four convolution layers in parallel. These four convolution layers have the same kernel size of 3 and different dilated rates of [0,1,2,3]. These layers can extract the information from the feature map with different sampling rates, which considers objects of different sizes. By using convolutions with stride two, these four layers shrink the output feature map by a factor of 4 while preserving the spatial resolution. Each layer has 1/4 of the desired output channels, and the outputs from these layers are concatenated. After that, a final convolution layer and a batch normalization layer are used in the block. If the input before a multi-scale pooling block has a size of $H \times W \times C$, the output feature map has a size of $(H/2) \times (W/2) \times C$. The output dimensions match the output from an ordinary max-pooling layer with stride 2. However, the block preserves spatial resolution and extracts features for objects of different sizes without generating feature maps at different scales. As a result, the memory and computation costs are reduced. For comparison, atrous convolution used in DeepLab also has a benefit of spatial resolution preservation, but the output feature map has a dimension of $H \times W \times C$, which is 4 times larger than using the multi-scale pooling block.[16]

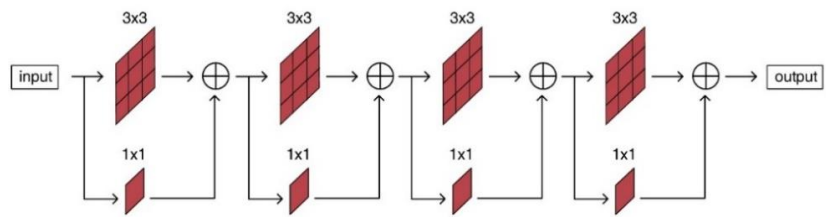

Figure 4. The Residual Forward Path

- The Residual Forwarding Path (Figure 4) is adapted from Multi-Residual U-net.[21] Unlike the original U-net architecture, the feature maps generated during the down-sampling process are not passed forward directly to the up-sampling side. By employing the residual path, non-linear operations are introduced to reduce the semantic gap between encoder and decoder.[21]

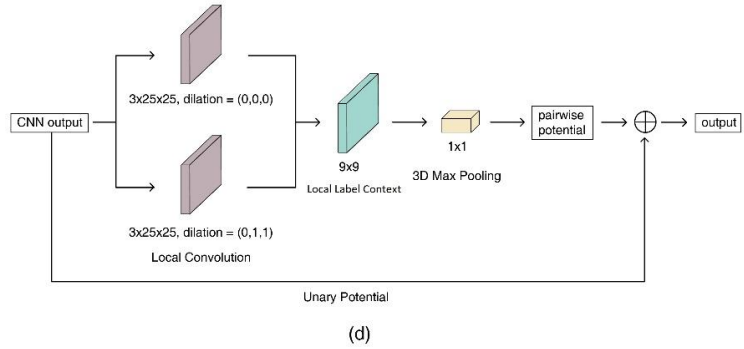

(d)

Figure 5. The MRF Block

- The MRF Block (Figure 5) that we designed was inspired by Deep Parsing Network.[17] The output of CNN is used as the unary potential of the MRF, and the MRF block uses local convolution layers, a global convolution layer, and a 3D max-pooling layer to calculate the pairwise potential. By minimizing the energy function of the MRF, CNN and MRF are trained jointly. A $50 \times 50 \times 3 \times 1$ filter (with 7500 parameters) is used for local convolution in Deep Parsing Network.[17] The structure has been revised to use two $25 \times 25 \times 3 \times 1$ filters (with 1875 parameters each) with different dilation rates in parallel. This modification reduces half of the parameters and performs local convolution at different scales.

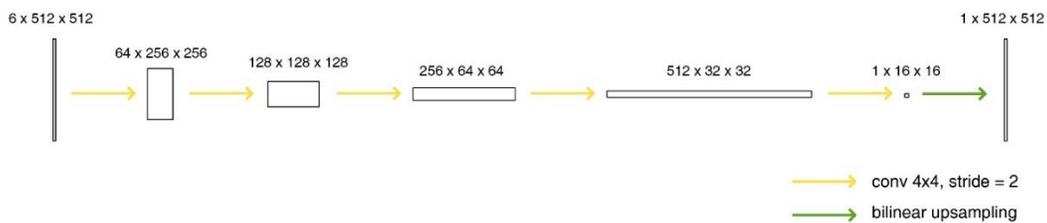

Figure 6. Discriminator Network (D-net)

The D-net consists of four down-sampling convolution layers, one classification layer, and bilinear up-sampling layers (Figure 6). The input to the discriminator is the product of the original CT $x_n$ and a predicted segmentation $\widetilde{y_n} = S(x_n)$ or the ground-truth label $y_n$. The D-net is trained to label $D(x_n \cdot y_n)$ as 1 (real) and $D(x_n \cdot \widetilde{y_n})$ as 0 (fake) for each pixel. The S-net is trained to generate a plausible $\widetilde{y_n}$ that can "fool" the D-net to label it as 1 (real) for each pixel.

**2.3 Adaptive Loss Function**

We adopted the loss function proposed in Luc's work[20] for the adversarial joint training:

$$\ell(\boldsymbol{\theta}_s, \boldsymbol{\theta}_d) = \sum_{n=1}^{N} \ell_{mce}(S(x_n), y_n) - \lambda[\ell_{bce}(D(x_n \cdot y_n), 1) + \ell_{bce}(D(x_n \cdot \widetilde{y_n}), 0)], \quad (1)$$

$$\ell_{mce}(\hat{y}, y) = -\sum_{i=1}^{H \times W} \sum_{c=1}^{C} y_{ic} ln \hat{y}_{ic}, \quad (2)$$

$$\ell_{bce}(\hat{z}, z) = -\sum_{i=1}^{H \times W} [z_i ln \hat{z}_i + (1 - z_i) ln(1 - \hat{z}_i)], \quad (3)$$

where $\boldsymbol{\theta}_s$ and $\boldsymbol{\theta}_d$ are the parameters of the S-net and D-net, respectively, H and W the height and width of the input image, $C$ denotes the number of classes, $N$ the number of training samples, $x_n$ represents one input CT image, $y_n$ denotes the corresponding ground truth segmentation map, and $\widetilde{y_n}$ the prediction generated by the network. Equation 2 is the multi-class cross-entropy loss for prediction $\hat{y}$, which is the negative log-likelihood of the ground-truth map $y$ with the one-hot encoding.[20] Similarly, Equation 3 is the binary cross-entropy loss for prediction $\hat{z}$ and ground truth $z$. The loss is minimized with respect to $\boldsymbol{\theta}_s$ but maximized with respect to $\boldsymbol{\theta}_d$.[20]

The loss function of the segmentation network is:

$$\mathcal{L}_S = \sum_{n=1}^{N} \ell_{mce}(S(x_n), y_n) - \lambda \ell_{bce}(D(x_n \cdot \widetilde{y_n}), 0), \quad (4)$$

where $S(\cdot)$ and $D(\cdot)$ denote the S-net and D-net, respectively, and $\lambda$ is the hyper-parameter for balancing the training of two networks.

The loss function is the difference between the multi-class cross-entropy loss of the segmentation network and the binary cross-entropy of being caught by the discriminator. For the first part of segmentation network loss, an adaptive multi-class cross-entropy loss is introduced to replace the original multi-class cross-entropy:

$$\ell_{mce}^* = -\sum_{i=1}^{H \times W} \sum_{c=1}^{C} w_c y_{ic} ln \hat{y}_{ic}. \quad (5)$$

In the novel loss function, the loss generated from each class is weighted by a coefficient, which is calculated based on the size of the organ and the current accuracy after each validation step. In CT scans, organs can have very different sizes in different CT slices. Larger organs have more pixels on the scan, and they have higher weights in determining the total loss with the original $\ell_{mce}$. In a pelvic CT, the prostate and rectum have much smaller sizes than the bladder. Besides, some organs, such as left and right femurs, are relatively easy to contour due to their higher contrast and sharper boundary. The new loss function is introduced to address the two drawbacks of the unweighted loss function. The weight in Equation 5 is increased for organs with smaller sizes that are hard-to-contour. The formula for the weight is:

$$w_i = 2 - DSC_i + \ln \frac{number\ of\ pixel\ of\ all\ classes}{number\ of\ pixel\ of\ class\ i}. \tag{6}$$

where $DSC_i$ is calculated by the last validation step, and the second term is calculated batch-wise based on the ground-truth label map. The loss function for the discriminator network is:

$$\mathcal{L}_D = \sum_{n=1}^{N} \ell_{bce}(D(\pmb{x}_n \cdot \pmb{y}_n),1) + \ell_{bce}(D(\pmb{x}_n \cdot \widetilde{\pmb{y}_n}),0) \tag{7}$$

**2.4. Training the Model**

The proposed model was trained on two RTX 2080 Ti GPUs in parallel to expedite the training. We implemented the new model with PyTorch and used Adam optimizer and the polynomial learning rate scheduler.[23] The training process of ARPM-net has three stages: (1) coarse training with the initial learning rate of 1e-2 for 10~20k iterations, (2) fine-tuning with the initial learning rate of 1e-4 for 20~30k iterations, and (3) adversarial training with the discriminator for 20~40k iterations. The adversarial training requires some delicate hyper-parameter tunning.[24] The steps of one adversarial training iteration include a) feeding the input forward through the segmentation network to compute the multi-class cross-entropy loss $\ell_{mce}(S(\pmb{x}_n),\pmb{y}_n)$; b) computing the input into the discriminator network and feeding the resulting input forward through the discriminator network to compute the loss $\mathcal{L}_D = \sum_{n=1}^{N} \ell_{bce}(D(\pmb{x}_n \cdot \pmb{y}_n),1) + \ell_{bce}(D(\pmb{x}_n \cdot \widetilde{\pmb{y}_n}),0)$; c) computing the loss for segmentation network $\mathcal{L}_S = \sum_{n=1}^{N} \ell_{mce}(S(\pmb{x}_n),\pmb{y}_n) - \lambda \ell_{bce}(D(\pmb{x}_n \cdot \widetilde{\pmb{y}_n}),0)$; and d) backpropagating $\mathcal{L}_D$ through the discriminator network and backpropagating

$\mathcal{L}_S$ through the segmentation network. Using adversarial training for fine-tuning will further improve the performance of the model. The average processing time (including the forward and backward propagation) during the training phase is roughly 0.5sec/image and the prediction time during the evaluation phase is about 0.3sec/image. The whole network takes roughly 8 hours to complete all training stages.

## 3. RESULTS

The ARPM-net method performed multi-organ segmentation on pelvic CT images and generated accurate contours on the test dataset (Figure 7). It produced accurate contours of OARs, including the prostate, bladder, rectum, left femur, and right femur, within 0.3 seconds.

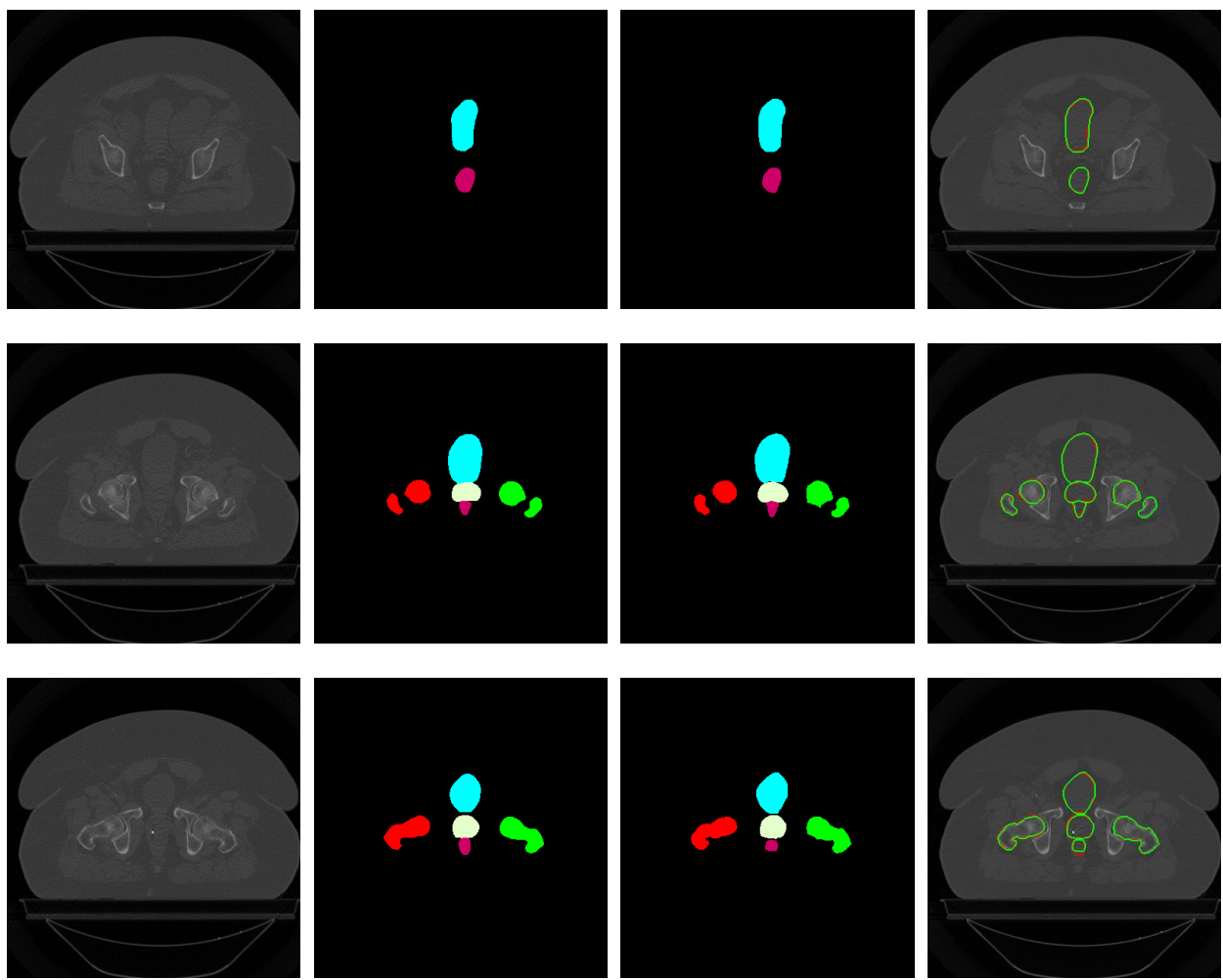

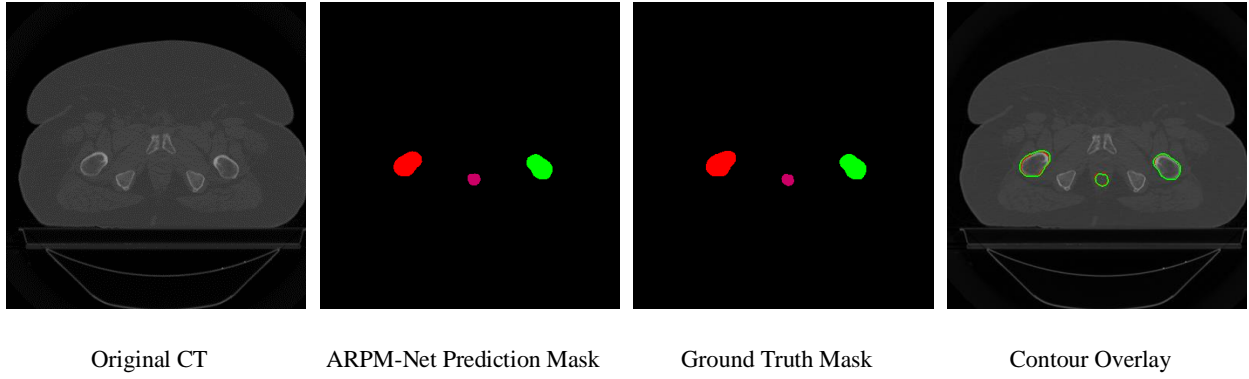

| Original CT | ARPM-Net Prediction Mask | Ground Truth Mask | Contour Overlay |

Figure 7. Segmentation results from the ARPM-net method. In the contour overlay column, green contours are the ground-truth labels, and red contours are predictions from ARPM-net. Shown in the figure are the CT images and contours for the same test case at different slices, ordered from superior to inferior.

On the CT images that we analyzed, the segmentations predicted by the ARPM-net (the second column of Figure 7) closely resembled the ground-truth masks (the third column of Figure 7). The contour overlay (the fourth column of Figure 7) also indicated that ARPM-net generated contours that closely resembled manual contours. The performance was evaluated by four widely used metrics for evaluating medical segmentation: the Dice Similarity Coefficient (DSC), Average Hausdorff Distance (AHD), Average Surface Hausdorff Distance (ASHD), and relative Volume Difference (VD).[4,14,25] The DSC metric is defined as:

$$DSC = \frac{2|True \cap Pred|}{|True| + |Pred|}, \qquad (8)$$

where $|True|$ and $|Pred|$ stand for the numbers of voxels in the ground-truth and predicted masks.

The Hausdorff Distance (HD) is defined as the maximum distance of a set to the closest point in the other set.[26] Mathematically, the Hausdorff Distance[26] from set $X$ to set $Y$ is defined as:

$$HD(X,Y) = \max_{x \in X}(\min_{y \in Y}(d(x,y))), \qquad (9)$$

where $d(x,y)$ is the Euclidean distance between point $x$ in voxel set $X$ and point $y$ in voxel set $Y$.[14] The AHD[14,25] between ground truth contour $X$ and segmentation result contour $Y$ is defined as the maximum of the point-wise average distance from points in X to the nearest point in Y and the point-wise average distance from points in Y to the nearest point in X:

$$AHD(X,Y) = max(\frac{1}{|X|}\sum_{x \in X} \min_{y \in Y} d(x,y), \frac{1}{|Y|}\sum_{y \in Y} \min_{x \in X} d(y,x)), \quad (10)$$

where X and Y denote the voxel set of ground truth and segmentation result, respectively; and $d(x,y)$ is the Euclidean distance from point x in set X to point y in set Y; and $d(y,x)$ is the Euclidean distance from point y in set Y to point x in set X.[25] The Average Surface Hausdorff Distance (ASHD)[14,27] is defined as the symmetrical point-wise average distance between points on one surface X to the nearest point on the other surface Y:

$$ASHD(X,Y) = \frac{1}{2}(\frac{1}{|X|}\sum_{x \in X} \min_{y \in Y} d(x,y) + \frac{1}{|Y|}\sum_{y \in Y} \min_{x \in X} d(y,x)), \quad (11)$$

where X and Y denote the voxel set of ground truth and segmentation result surface, respectively; and $d(x,y)$ is the Euclidean distance from point x in set X to point y in set Y, and $d(y,x)$ is the Euclidean distance function from point y in set Y to point x in set X.[27] The relative Volume Difference (VD)[28] is used to evaluate the tendency of under-segmentation or over-segmentation, which is defined as:

$$VD(R,G) = \frac{|R|-|G|}{|G|} \cdot 100\%, \quad (12)$$

where |R| and |G| stand for the volume of segmentation results and ground truth, respectively.

|  |  | DSC (±SD) | AHD (mm±SD) | ASHD (mm±SD) | VD (%) | *p-value* |
|---|---|---|---|---|---|---|
| Prostate | **Baseline 1** | 0.83(±0.12) | 1.79(±2.01) | 2.85(±2.46) | **- 5.71** | 0.010* |
|  | **Baseline 2** | 0.84(±0.15) | 1.82(±1.78) | **2.11(±2.27)** | - 8.56 | 0.017* |
|  | **ARPM-net** | **0.88(±0.09)** | **1.48(±1.58)** | 2.21(±1.73) | - 6.65 | - |
| Bladder | **Baseline 1** | 0.95(±0.09) | 2.12(±1.65) | 3.21(±2.92) | + 2.33 | 0.077 |
|  | **Baseline 2** | 0.95(±0.07) | 1.93(±2.54) | 2.80(±2.12) | + 1.36 | 0.099 |
|  | **ARPM-net** | **0.97(±0.08)** | **1.90(±1.75)** | **2.09(±1.85)** | **- 1.19** | - |
| Rectum | **Baseline 1** | 0.84(±0.10) | 2.36(±1.51) | 3.49(±3.74) | **+ 4.94** | 0.028* |
|  | **Baseline 2** | 0.85(±0.15) | 2.84(±2.48) | 3.31(±2.92) | + 7.82 | 0.061 |
|  | **ARPM-net** | **0.86(±0.06)** | **2.21(±1.65)** | **3.21(±2.82)** | + 6.47 | - |
| Femur_L | **Baseline 1** | 0.96(±0.02) | 1.98(±1.39) | **1.82(±1.13)** | - 1.74 | 0.120 |
|  | **Baseline 2** | 0.96(±0.01) | 2.05(±1.43) | 1.97(±1.22) | + 2.69 | 0.151 |
|         | ARPM-net   | 0.97(±0.01)  | 1.85(±1.56)  | 1.92(±1.35)  | + 1.15 | -     |
| Femur_R | Baseline 1 | 0.97(±0.01)  | 1.89(±1.42)  | 2.16(±1.60)  | + 3.26 | 0.171 |
|         | Baseline 2 | 0.97(±0.03)  | 1.93(±1.29)  | 2.05(±1.06)  | + 4.73 | 0.182 |
|         | ARPM-net   | 0.97(±0.01)  | 1.88(±1.32)  | 1.98(±1.75)  | + 1.55 | -     |

Table I. Comparative analysis of the 10-fold cross-validation performance of ARPM-net and baseline methods for semantic segmentation of multiple organs in pelvic CT images. Baseline 1 is the residual U-net[29], and baseline 2 is the multi-residual U-net[21]. The best scores among the three comparing methods for each organ are in red. In the last column, "*p-value" was* calculated by comparing each baseline method with the new ARPM-net method, and we used * for statistically significant improvements (*p-value* < 0.05).

Table I shows the 10-fold cross-validation performance of our ARPM-net and the baseline methods. The accuracy of each of the femurs was close to 100% thanks to their higher contrast levels and sharper boundaries. In comparison, it is more challenging to accurately contour the prostate and rectum because of their smaller sizes and lower contrast. Nevertheless, the new ARPM-net method generated high-quality contours for all these organs. The results of AHD and ASHD show that contours generated by the ARPM-net closely resemble the ground truth, and the results of VD show that the model has tendencies to over-segment the rectum (+6.47%) and femurs (left: +1.15%, right: +1.55%), and to under-segment the prostate (-6.65%) and bladder (-1.19%). Compared to a manual segmentation process that took 20-30 min/patient for the prostate alone, ARPM-net produced contours of five organs within 2 min/patient on a machine with only two GPUs. These results showed that the new method could improve the quality of routine clinical practice by providing highly consistent contours with high quality and efficiency, which validated our first hypothesis.

The second hypothesis that our proposed method can outperform the baselines can be tested by the results in Table I. All three methods achieved high accuracy on the left and right femur. As shown in Figure 8, the two baseline methods failed to deliver accurate segmentations on small and low-contrast organs (prostate and rectum). The baseline 2 (Multi-residual U-net[21]) performed slightly better than baseline 1 (2D U-net[8]) for segmenting the prostate and rectum, and our new method outperformed both baselines. We computed the *p-value* of five OARs comparing our method with both baselines (Table I). Our method significantly improved the segmentation accuracy of hard-to-contour organs (prostate and rectum) and outperformed both baselines. As stated in section 2.1, we also tested our ARPM-net with 20 test cases.

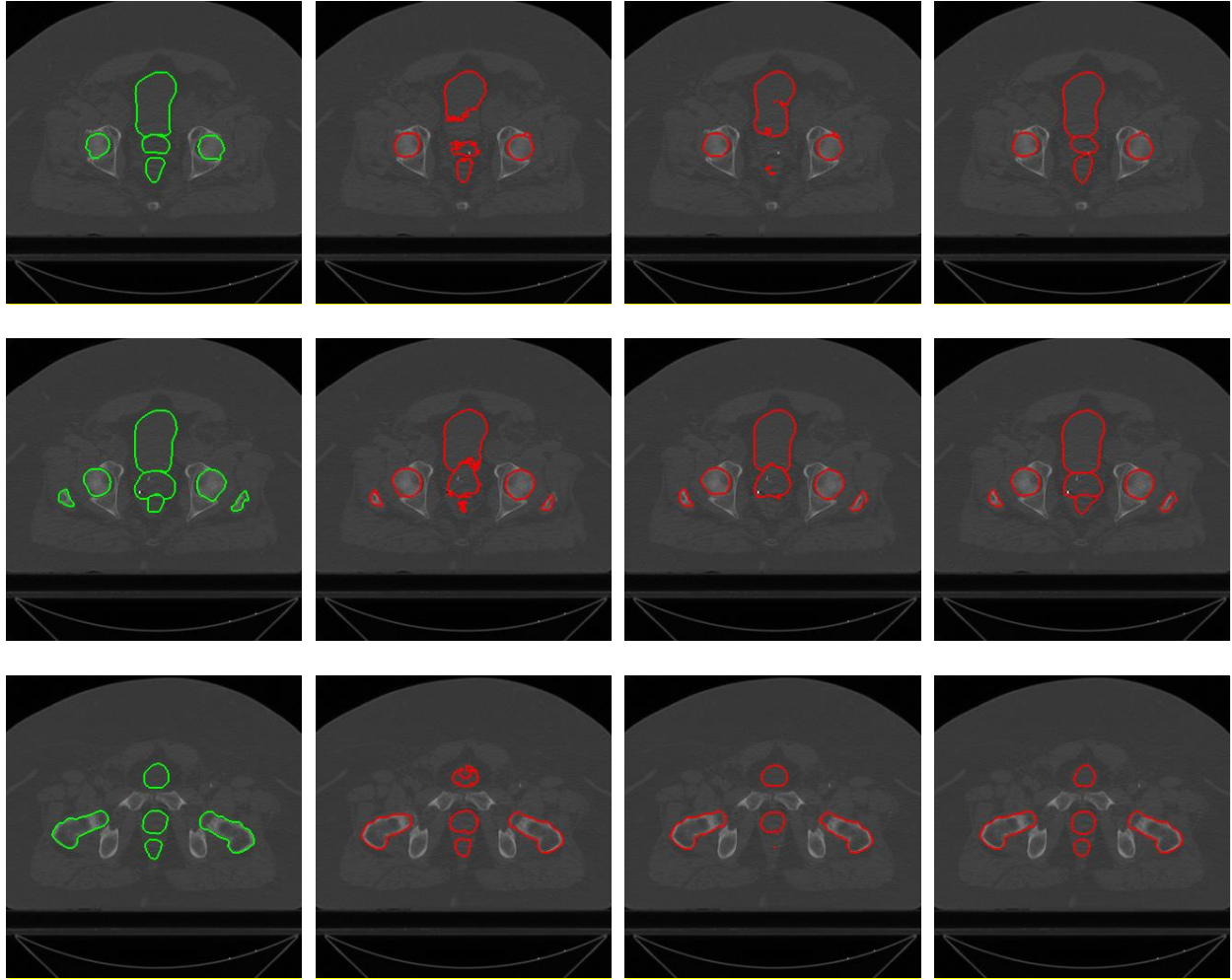

| Ground-truth Overlay | Baseline 1 Contour Overlay | Baseline 2 Contour Overlay | ARPM-Net Contour Overlay |

Figure 8. Segmentation results comparing the baselines with ARPM-net. Green contours are the ground-truth labels, and red contours are predictions. Shown in the figure are the CT images and contours for the same test case at different slices, ordered from superior to inferior.

## 5. DISCUSSION

Despite that a direct comparison is not possible, Table II shows the comparison between the performance of our proposed method and the corresponding metrics of other state-of-the-art methods in the literature. The geometrical shape model[5], multi-atlas based segmentation[30], and boundary regression[31] methods are not deep learning-based. The deep multi-task random forest[7] and deep decision forest[6] are combinations of traditional machine learning techniques and deep learning implementation. Besides the non-deep-learning and hybrid deep learning approaches, we also compared ARPM-net with deep learning approaches, including deep dilated CNN[32], 2D U-net[14], and 3D U-net[4] using the same patient dataset. The deep dilated CNN[32], as well as all the above non-deep-learning and hybrid deep learning approaches, are full pelvic CT segmentation[5-7,30,31]. The 2D U-net and 3D U-net are ROI segmentation methods, which generate contours for different organs separately on different patches, and the latter one uses a 2D U-net for patch localization and a 3D U-net for ROI segmentation[4].

| Model | Prostate DSC (±SD) AHD (mm±SD) ASHD (mm±SD) | Bladder DSC (±SD) AHD (mm±SD) ASHD (mm±SD) | Rectum DSC (±SD) AHD (mm±SD) ASHD (mm±SD) | Femur_L DSC (±SD) AHD (mm±SD) ASHD (mm±SD) | Femur_R DSC (±SD) AHD (mm±SD) ASHD (mm±SD) |
|---|---|---|---|---|---|
| **Geometrical Shape Model[5] (2014)** | 0.87(±0.001) 9.98(±3.4) - | 0.89(±0.001) 25.1(±4.6) - | 0.82(±0.001) 13.5(±5.1) - | N/A | N/A |
| **Multi-atlas Based Segmentation[30] (2014)** | 0.85(±0.004) - - | 0.92(±0.002) - - | 0.80(±0.007) - - | N/A | N/A |
| **Boundary Regression[31] (2015)** | 0.88(±0.0002) - - | N/A | 0.84(±0.001) - - | N/A | N/A |
| **Deep Multi-task Random Forest[7] (2016)** | 0.87(±0.004) - - | 0.92(±0.005) - - | 0.88(±0.005) - - | N/A | N/A |
| **Deep Dilated CNN[32] (2017)** | 0.88 - - | 0.93 - - | 0.62 - - | 0.92 - - | 0.92 - - |
| **Deep Decision Forest[6] (2018)** | 0.75-0.76 - - | 0.94-0.97 - - | 0.71-0.82 - - | 0.96-0.97 - - | 0.96-0.97 - - |
| **2D U-Net[14] (2018)** | 0.88(±0.12) 0.4(±0.7) 1.2(±0.9) | 0.95(±0.04) 0.4(±0.6) 1.1(±0.8) | 0.92(±0.06) 0.2(±0.3) 0.8(±0.6) | N/A | N/A |
| **2D U-Net Localization, 3D U-Net Segmentation[4] (2018)** | 0.90(±0.02) 5.3(±2.8) 0.7(±0.5) | 0.95(±0.02) 17.0(±14.6) 0.5(±0.7) | 0.84(±0.04) 4.9(±3.9) 0.8(±0.7) | 0.96(±0.03) - - | 0.95(±0.01) - - |
| **ARPM-Net (Our Method)** | 0.88(±0.11) 1.58(±1.77) 2.11(±2.03) | 0.97(±0.07) 1.91(±1.29) 2.36(±2.43) | 0.86(±0.12) 3.14(±2.39) 3.05(±2.11) | 0.97(±0.01) 1.76(±1.57) 1.99(±1.66) | 0.97(±0.01) 1.92(±1.01) 2.00(±2.07) |

Table II. Comparative analysis of the performance of ARPM-net on the test set (20 cases) and eight state-of-the-art methods for semantic segmentation of multiple organs in pelvic CT images. "-" denotes that a certain metric is not available, and "N/A" means that certain organs are not segmented by the method.

One of the advantages of our new method is the high performance of all five classes spanning different organs with only one network architecture and one forward propagation. Larger organ (bladder) and smaller organs (prostate and rectum), high-contrast organs with sharp boundaries (left femur and right femur), and low-contrast organs with unclear boundaries (prostate) were contoured at the same time, and the training process was balanced by using our proposed adaptive loss function.

ARPM-net achieved state-of-the-art performance for multi-organ segmentation on full pelvic CT scans. ARPM-net had high average DSCs on multiple organs with low standard deviations. Segmentations generated by ARPM-net also had lower AHD to ground-truth compared to other listed CT segmentation methods. Although the DSC score was slightly lower than that reported by the ROI segmentation methods[4,14], our new ARPM-net method does not need a separate localization network or have to extract patches of ROIs.[4,14]

Despite its desirable features, the new method can be further improved. The first limitation of the new method is its delicate design and tuning process, which is typical for most methods using adversarial networks. The adversarial training process requires some careful hyperparameter tuning. The architecture of the discriminator network must be designed carefully. It is a common issue for adversarial training models to design two "players" to have compatible competitive capacities to improve upon each other's performance over iterations to reach equilibrium.[18] "Mode collapse" may occur if one of the two players is too weak or too strong because the training of the other player may be jeopardized.[33] In the case of semantic segmentation, for example, if the discriminator is not designed correctly, it may not be able to extract features from its input to distinguish fake data from real input. Consequently, the adversarial loss will contribute little to the training of the segmentation network. The segmentation network can always "fool" the discriminator network regardless of what output it generates, resulting in a "mode collapse", meaning that the segmentation network does not learn to perform quality segmentation, but rather learn to fool the discriminator instead.[18] The second limitation of the work is that we used a small test set with only 20 cases, considering the variety of pelvic CT images. The proposed method will be further validated on a larger dataset to show the generalizability of the algorithm. The third limitation of our new method is that it did not utilize the information between adjacent slices. The shape of OARs should have coherent shapes on adjacent slices, and such information can be exploited if we use 3D models. We will extend and improve our ARPM-net to perform 3D volume segmentation in our future work.

## 6. CONCLUSION

In this work, we developed a CNN based method for semantic segmentation of pelvic CT images. The novel ARPM-net method integrated Markov Random Field with deep learning frameworks efficiently and used adversarial training to utilize style-wise loss for fine-tuning. ARPM-net adapted the multi-residual U-net structure and multi-scale pooling blocks to enhance the original max-pooling layers. The new multi-scale pooling block effectively reserved the spatial resolution and extracted features at multiple scales during the down-sampling process. It also reduced the memory requirement for training by generating smaller feature maps compared to atrous convolution layers. We improved the MRF block from the DPN model by performing local convolution at multiple scales, which significantly reduced the number of parameters.

ARPM-net achieved state-of-the-art performance on the male pelvic CT dataset, as evaluated by 10-fold cross-validation and a separate test set. By proposing a novel weighted loss function, we overcame the obstacle of training small and low-contrast organs in a multi-organ segmentation model. ARPM-net outperformed the other full CT segmentation methods. It reduced computational cost and memory requirements and achieved a competitive performance comparing to the ROI segmentation models. The low memory requirement and fast speed make the new method efficient and easy to be adopted in routine clinical practice.

**ACKNOWLEDGMENTS**

We would like to thank Varian Medical Systems for their financial support through a research grant.